\documentclass[11pt]{article}

\usepackage{pdfpages}
\usepackage{multirow,multicol,soul}
\usepackage[normalem]{ulem}
\usepackage{tabularx}
\usepackage{longtable,eurosym}
\usepackage{booktabs}
\usepackage{amsmath,bbm}
\usepackage{amssymb, epsfig}
\usepackage{float,graphicx,subfigure}
\usepackage[utf8]{inputenc}
\usepackage[counterclockwise]{rotating}
\usepackage{lscape,setspace}
\usepackage[top=2cm, bottom=2cm, left=2cm , right=2cm]{geometry}
\usepackage{url,hyperref}
\usepackage{color}
\usepackage[T1]{fontenc}
\usepackage[square,sort,comma,numbers]{natbib}
\setcitestyle{authoryear,open={(},close={)}}

\definecolor{darkgreen}{RGB}{51,153,55}

\usepackage{todonotes}

\graphicspath{{Plots/}}

\date{}

\begin{document}

\title{Managing Cyber Risk, a Science in the Making}

\author{Michel Dacorogna\textsuperscript{a} and Marie Kratz\textsuperscript{b}
\\[1ex]
\small
\textsuperscript{a} PRIME RE Solutions, Zug, Switzerland; Email: michel.dacorogna@prs-zug.com\\
\small
\textsuperscript{b} ESSEC Business School, CREAR, Cergy, France; Email: kratz@essec.edu
}

\maketitle

\begin{abstract}
\noindent Not a day goes by without news about a cyber attack. Fear spreads out and lots of wrong ideas circulate. This survey aims at showing how all these uncertainties about cyber can be transformed into manageable risk. After reviewing the main characteristics of cyber risk, we consider the three layers of cyber space: hardware, software and psycho-cognitive layer. We ask ourselves how is this risk different from others, how modelling has been tackled and needs to evolve, and what are the multi-facetted aspects of cyber risk management. This wide exploration pictures a science in the making and points out the questions to be solved for building a resilient society.
\end{abstract}

\\[2ex]


\newpage
\tableofcontents
\newpage
\section{Introduction}
\label{sec-intro}
\vspace{-2ex}

As of late, there is no need anymore to emphasize the importance of managing cyber risk. Both the lock-down due to COVID and the war in Ukraine have shown that our reliance on IT solutions to communicate, work and fight puts us in danger when determined hackers target the IT systems. Numerous examples of these crowd the front pages of newspapers. It suffices to remember last Russian attacks on Ka-Sat at the dawn of the invasion of Ukraine. They targeted a telecommunications company used both by the Ukrainian army and many private customers in other parts of Europe, notably France and Italy. As the pressure mounts on society to become more resilient to cyber attacks, it also mounts on insurance companies to provide appropriate cyber covers to the market. Thus, as actuaries, we need to address the issue rapidly and come up with innovative solutions to find good data representing this risk and to propose appropriate models for valuing it, as well as efficient hedging strategies.

Many presentations on cyber risk start by pointing out the difficulties in modelling it because of the rapidly changing IT landscape, based on the constant development of computing and communication technologies: What we thought true yesterday is no longer true today. As much as this observation is true, it should not prevent us from looking for solutions to the problem. A great Russian physicist, Lev Landau, used to tell his PhD students that science was about finding constants in Nature. It is our role as scientists to come with models that can predict with a certain accuracy the degree of risk presented by cyber attacks. The ambition of this review paper is to show that there are many paths to come up to a successful understanding of this risk and that we should engage in big efforts to do so. It is the condition for rendering society more resilient to it and for pushing the insurance market to provide adequate covers to this risk. We already have some hints that it should be insurable, with some exclusions, as usual in an insurance policy. However, the first step is to gain a good understanding of the peculiarities of this risk.

\begin{figure}[h]
  \centering
  \includegraphics[scale=0.5]{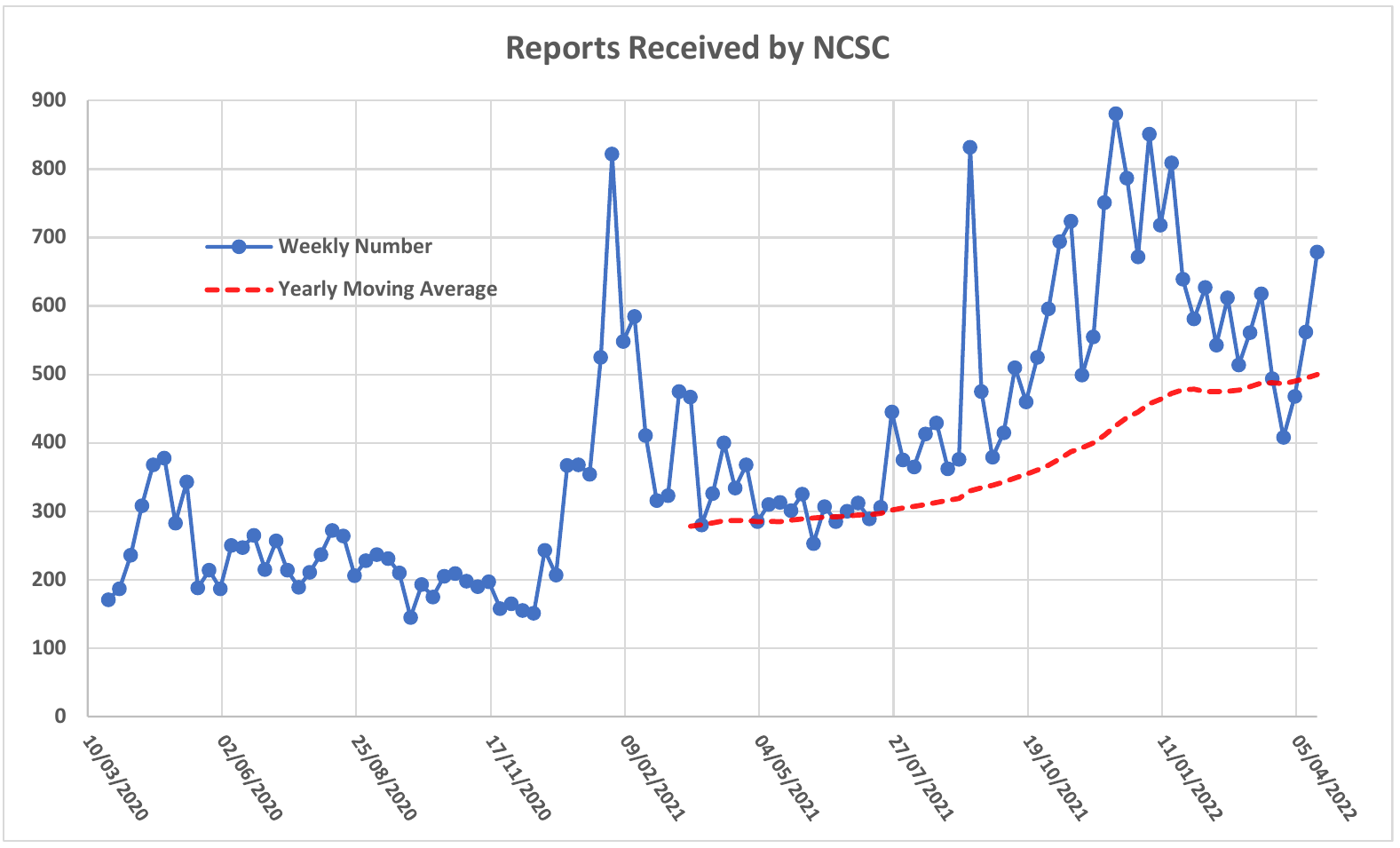}
  \caption{ \label{fig:NCSC} \sf The weekly incidents reported to the Swiss NCSC and its yearly moving average.}
\end{figure}

The number of reported attacks is in the rise. As an evidence of this, we plot, for instance, in Figure~\ref{fig:NCSC} the numbers of cyberincidents currently reported by the public and SMEs to the Swiss National Cybersecurity Centre (NCSC) as well as the yearly moving average. We see that, over a year, the average of weekly numbers has grown from a little less than 300 a week to more than 500 a week, an increase of more than 66\%! This is representative of a general phenomenon we can observe in various countries (see, for instance~\citep{Ventre2020}). As early as 2017, 80\% of companies in France had been subject to at least one cyber attack\footnote{Source:  OpinionWay survey for CESIN (Club des Experts de la Sécurité de l’Information et du Numérique) – Jan. 2018} and, according to the French {\it Gendarmerie Nationale} (the French national police force, denoted GN), 80\% of cyber attacks concern Small and Medium Businesses. We see that everyone and every organization are targets and need to protect themselves against these intrusions.

Another important characteristic of cyber risk is a highly interconnected world. Think of the cloud services that are dominated by few providers like Amazon Web Services (AWS), or Azure from Microsoft, or Google Cloud. These three giants dominate with 64\% (33\% AWS, 22\% Azure, 9\% Google cloud) market share the global cloud services in the last quarter of 2021\footnote{Source: Statista (\url{www.statista.com/statistics/967365/worldwide-cloud-infrastructure-services-market-share-vendor/})}. Such a concentration constitutes a risk shared by all customers, which is the very definition of systematic risk. The wide use of the same operating system like MS Windows or Android represents another of these systematic risks. The connection to the Internet is in itself another part of the interdependence of IT systems. Any system linked to others is susceptible of an attack. As some cyber security experts say, the only way to protect oneself completely is to stay disconnected. This is of course not possible anymore without losing access to an enormous amount of resources. So, our approach to model this risk must take into account this highly connected world.

Understanding cyber risk requires having data to find empirical properties and be able to develop good models. Paradoxically, IT systems produce a huge amount of data that has given birth to what is called today "data science", but there is still a lack of data to analyze the cyber risk itself. There are many reasons for this. One of them is the fact that victims are often reluctant to report the attacks as it could give them bad publicity. This is changing as new regulations are put in place that require companies to inform about attacks and to engage in forensic activities to understand how the attack was possible. On one had, the lack of insurance data due to the fact that cyber insurance is a relatively new product, nn the other hand, insurances are still reluctant to offer extensive covers that would produce large claims. Nevertheless, data exist and are used by various research groups to come up with models. There are data coming from various sources like a database on operational incidents in financial institutions~\citep{Eling2016}, data produced by telecommunication firms~\citep{Dalmoro2020}, or complaints filed at the French Gendarmerie Nationale~\citep{Ventre2020}. \cite{Zaengerle2022} give a useful list of available data in their Table~1. Moreover, insurance companies, although limited in coverage due to their reluctance to sell such products, have their own data, which they jealously guard for their own actuaries. In the long run, they will open up their databases because it is in their interest to compare their own data to that of their competitors and to reach a common understanding of cyber risk.

Related to the search for data is the question of the approach to modelling. Should we try to model the causes of cyber attack? In this case, we need data on the systems, the networks, the assets of interest to the hackers. Should we, to the contrary, model the consequences of IT system failures due to hackers? In this case, we need data on business interruption, on recovery capacities. Do we want to model the systemic risk? Then we need to look for network configuration data or to stress scenarios. Do we want to model probability distributions for pricing insurance covers? In this case, claims data would be the basis to obtain a price based on experience. Of course, actuaries also use other forms of modelling based on exposures. In this case, they will look for data about the client IT systems, their domain of activity, their geographical location, ... We see that there are many ways to collect information to build models.

In this paper, we consider cyber risk as a whole, from its characteristics to its management, through its modelling,  in a multifaceted way questioning how cyber risk fundamentally differs from other risks or wether it is comparable to some of the risks we are already confronted with. By doing so, we would like to show how research on cyber risk is evolving and what are the main results obtained so far. In our work, three of the existing review papers on the topic will be helpful: \cite{Eling2020} concentrates on the business and actuarial science literature, \cite{Awiszus2022} on the modelling and pricing of cyber insurance, while \cite{Marotta2017} provides a good overwiew on the cyber insurance market and economics, as well as research challenges. Other information taken from the media or the specialized press, conferences, discussions with cyber securities and insurance companies or within the ASTIN cyber working group, will also feed our exploration of this problem. We present here also general concepts on cyber, to help draw possible ways to investigate further this complex risk. 

In Section~\ref{sec-CyberRisK}, we discuss the characteristics of this risk and introduce the concept of cyber space. Various approaches for modelling are presented in Section~\ref{sec-modelling}, while the consequences for risk management and insurance are discussed in Section~\ref{sec-riskmgmt}. Conclusions and perspectives are drawn in Section~\ref{sec-concl}.

\section{What is so special about cyber risk compared to other types of risk?}
\label{sec-CyberRisK}
\vspace{-2ex}

\subsection{Characteristics of cyber risk}

\begin{figure}[b]
   \centering
   \includegraphics[scale=0.7]{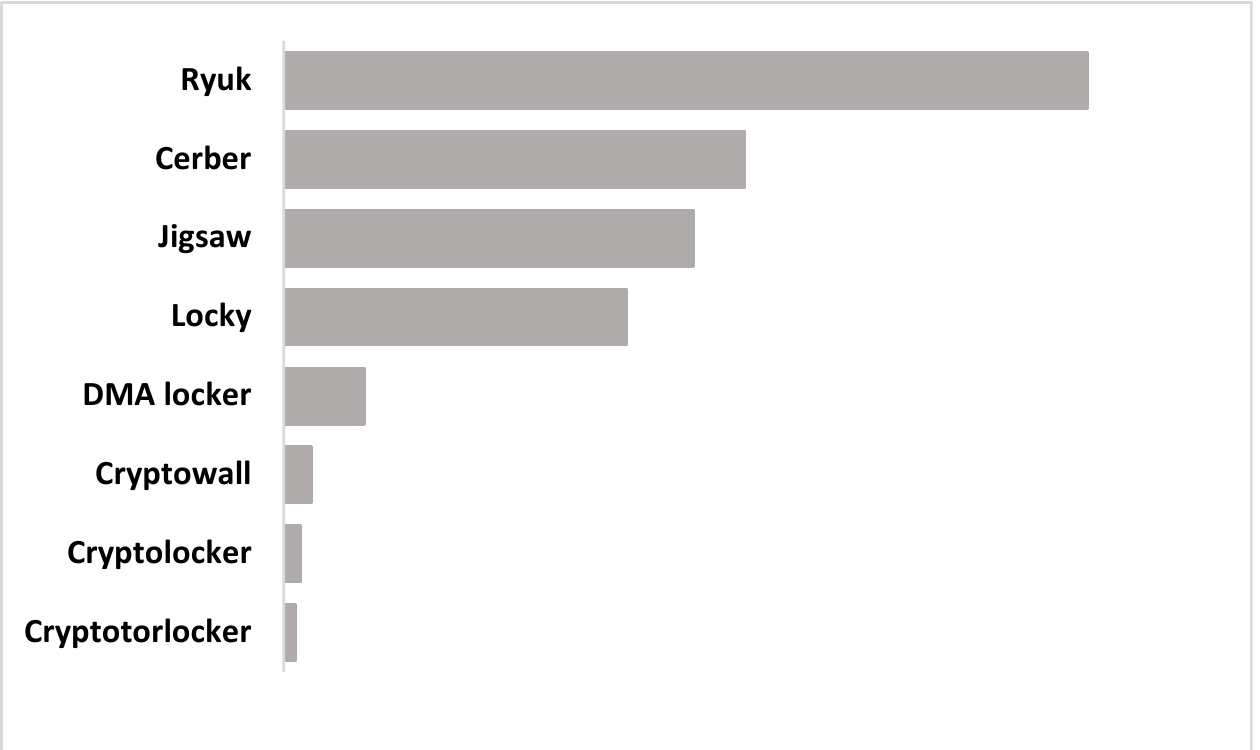}
   \parbox{280pt}{\caption{\label{fig:Alex} \sf \small The average ransom normalized by the time of activity for various ransomwares}}
\end{figure}

In the introduction, we mentioned the increase of the frequency of attacks. What is also apparent lately is an increase in the financial consequences of these attacks. For instance, according to a claim-study performed by Netdiligence (a company specialized in cyber risk readiness \& response services) on a private database containing 5,797 claims, the ransom demands sky rocketed from 59kUSD in 2016 to 751kUSD in 2021 (see Figure 1 in~\cite{Netdiligence2022}). This is only an illustration of a fact that appears in many reports written by consulting firms interviewing their clients. This also means that more and more financial resources are available for attackers. In an ongoing study (with A. Morin) on a database of ransoms we have access to, we can observe a similar behavior as just described. We illustrate it with the plot of the average ransom normalized by the time of activity; see  Figure~\ref{fig:Alex}. The ransoms paid by victims are followed through the cryptocurrency addresses given to them for paying the hackers. In the plot, we did not show the scale, nor the currency, as we just want to point out a clear distinction of the average amount between the four largest and the four smallest; the largest ones correspond to a more recent period (after 2016), while the smallest ones correspond to the period 2014-16 (except for cryptolocker on 2014-2018). It goes along with the findings of Netdiligence and others, that hackers better target their victims and are able to extort much more money.

It is why the level of sophistication of these attacks is increasing rapidly. It is the same old story of the sword and the shield. The competition between cyber security and hackers is going on at a high speed. This is one of the components that make cyber risk special, the high speed of changes in the risk landscape. Clearly, this is not unique to cyber, one experiences similar climate changes on a geological scale, but it remains one important characteristic of this risk. How confident can we be that the data we have collected so far actually represent the risks that we are going to face in the future? Thus, modelling cyber risk requires the use of both quantitative and qualitative assessment of the landscape. In particular, developing stress scenarios is an important technique for accurately assessing the highest risks.

A second important characteristic is, from the point of view of insurance and actuarial valuation, the fact that the targets of attacks are often intangibles, like data, reputation or political elections. The financial damages of those are very difficult to assess. They can be of vastly different sizes and would depend on subjective valuation by victims or insurance companies\footnote{We will encounter this difficulty later concerning the PRC database of number of data breaches without amount of money.}. No doubt that this fact could cause much litigation in courts and/or the need to rework the exclusion wording of insurance policies. Linked to this, the difficulty of localizing the origin of the attacks makes it difficult for companies to look for compensation and reduction of the claims.	 A new practice from insurers makes the cyber cover less attractive: They limit the possible payment for these intangible losses. Customers, particularly financial institutions, are asking for much larger covers than those the insurers are willing to sell. It is why, despite a strong demand, the offer does not match and the cyber insurance market does not grow at the speed required by society.

A third property that makes cyber risk special and complex is the potential of systemic failures due to attacks. We mentioned already in the introduction the highly connected world of IT systems. If one adds to this the crucial role they play today in many human activities, it becomes clear that we could be confronted to failures of dramatic consequences. The war in Ukraine has already shown that attacks aiming at a particular target, the communication system of the Ukrainian army, can have unintended spill-over on other unwanted targets, like the Italian or French communication system. Attacks against hospitals during the pandemic showed the vulnerability of our societies to ill-intentioned attackers. Until now, we have been able to limit the consequences of these attacks, but it is not certain that we  will always succeed in doing so. It is another important reason for the reluctance of insurance companies to be more active in this market.\\[1ex]

\subsection{Cyber space}
\label{ss-space}

The cyber risk affects all aspects of society, but develops in its own space. It is a virtual space but with concrete and physical impacts. Even though cyber space is virtual, it is organized similarly to the real space, with the same structure as in the real world, although built with new tools, namely information technology (IT) and artificial intelligence (AI). Special interest groups evolve within this space, with human strategies existing already for centuries but developed with innovative and far reaching technologies. No wonder that some hackers named a virus, Trojan horse, a malware appearing as a legitimate or appropriate software but compromising system security. 

So, we can include the cyber space as a fourth one to the list of our 3 well known worlds, terrestrial, sea and air. Nowadays, those four define our environment and interact strongly with each other. Cyber risk originates from this interaction. Computers, mobile phones, and today more and more the internet of things (IoT)\footnote{According to Strategy Analytics, by the end of 2021, more than 27 billion IoT and connected devices were deployed worldwide; see \url{www.strategyanalytics.com}} are the communication vectors between cyber and real spaces, making  cyber space pervasive in people's private and professional lives.   
As any risk, it has positive and negative sides, offering a new set of opportunities, but also new weapons that make us vulnerable. According to \cite{Ventre2016}, the cyber space is a set of three layers:
\vspace{-3ex}
\begin{enumerate}
\item A physical layer that corresponds to the IT infrastructure, including the web of cables that constitutes today the necessary vehicle for Internet, satellites, and cloud centers;
\vspace{-1ex}
\item A software layer that of course includes also malware;
\vspace{-1ex}
\item A psycho-cognitive layer that contains information, images, texts, sounds, social networks.
\end{enumerate}
\vspace{-2ex}
Thus, cyber risk can be situated in the three layers and should be addressed differently for each, taking also connections between them into account. For instance, companies are now hiding their servers so that they cannot be reached by bombs in case of war, while on the psycho-cognitive layer, very different risk management should be applied. Other examples concerns cyber security firms, who apply technologies, processes, and controls to protect systems, networks, programs, devices and data from cyber attacks. They address mostly the two first layers, but use the third one to collect data.

Many entities are acting in the cyber space: governmental agencies, states, interest groups, companies, associations, individuals on private and professional sphere, ... They exist in this space through a digital interface, as a website or email or mobile phone. It may also be important to take into account those various actors when interested in cyber risk modelling using game theory.
As in any space, risk is inherent to the organization and exchanges of those actors, reason why cyber security has been developed in parallel, professionally and academically. It is a top priority, specifically for governmental entities. The various actors can be spread out between the three layers mentioned above with different weights and possible impacts. Cyber security needs to address the three layers, with the risk of producing an over-control tool for populations, as already experimented in some countries. This is also a risk for democracy where social networks reinforce the spreading of fake news and, contrary to what we would have expected, favor the constitution of closed groups of individuals thinking alike, making it difficult or even impossible to build a consensus. It partly explains the raise of populism in Western democracies.  

Cyber crimes is a new category of crimes, targeting the same goal as usual crimes, but via new means. It requires attackers and defendants to adapt to cyber space. Evaluating the impact of a cyber crime is made difficult due to the nature of cyber weapons, less tangible and more pervasive, which impact can be detected in an unpredictable period of time (from very short to very long).
The same holds for cyber war between states or interest groups: It is based on old strategies but with new means; see e.g. \citep{Ventre2016}. Nevertheless, defining the notion of cyber war is less than obvious, because of the difficult task of identifying the origin of an attack and evaluating its damages; see e.g. \citep{Marsh2022} for a discussion on the topic. Those two conditions are central for defining war exclusions in legal and insurance terms. Indeed, (re)insurance companies are authorized to exclude (traditional) wars from their policies. It explains why the latter are working intensively on this topic, trying to come up with an explicit definition of cyber war, or at least of cyber war exclusions. Several propositions have been made, among which recent ones by Llyods (LMA 5564 to 5567\footnote{Source: \url{www.lmalloyds.com/LMA/News/LMA_bulletins/LMA_Bulletins/LMA21-042-PD.aspx}}), Munich-Marsh-AON  'war and cyber operation exclusion'\footnote{Source: \url{www.marsh.com/us/services/cyber-risk/insights/moving-towards-clarity.html}}, or Beazley\footnote{Source: \url{www.beazley.com/en-us/articles/addressing-catastrophic-cyber-risks}}. Cyber war, as any war, is the manifestation of one of the drivers for mankind, which always existed: power play between political, economical and military powers. Behind any attack, there is always a conflict of powers; thus, it is important to identify which one is concerned. There is always a technology to wage a war. While the life cycle of technology is quite short, political strategies remain remarkably similar over time.

\section{What is so special about cyber risk models compared to other types of risk models?}
\label{sec-modelling}
\vspace{-2ex}

\subsection{Framing the problem}
\label{ss:frame}
\vspace{-2ex}
%
Cyber risk has specific characteristics, as seen in the previous section.
First, cyber risk 
must be clearly defined, so that we know which aspect(s) is(are) taken into account when turning to modelling. There have been many attempts to come up with a good taxonomy for risks defining cyber. Once definitions are put in place and well accepted, it motivates and helps to collect data of incidents related to cyber. Another advantage of precise definitions is that they allow to measure the value of potential damages and manage the risk exposure in line with the companies’ or individuals’ risk appetite.

In terms of modelling, traditional actuarial techniques (frequency/severity) are difficult to apply for rating and risk accumulation control. This is due to both the short history of available data and the need to go beyond the pure loss history, as exposures are changing rapidly and new threats are constantly emerging. 
One way to consider this fast evolving environment is to develop stress scenarios considering {\it footprint} of particular events. They could be used, on one hand, to check if the capital of the insurer can handle such a scenario, on the other hand, to see if the probability of extremes evaluated through the model corresponds to that of the stress scenario. If the known probability of the stress scenario is higher than the probability of the loss scenario obtained with the model, we can then conclude that the model under-estimates extreme risk. Scenarios are very useful for model validation, or, if not enough data are available, to calibrate tail distributions on extremes imagined via scenarios (see \cite{Dacorogna2017}).
For accumulations, new conceptual thinking is needed, in particular in terms of supply chains, but also in terms of possible events with impact on the psycho-cognitive layer (e.g. reputational damages), which will, in turn, affect various insurance contracts.

Due to the interconnectedness and the strong suspiscion of the presence of systemic risks, the analysis of extreme risks is key for modelling cyber risk.
Quoting H. Rootz\'en during one of his conferences:\\[-6ex]
\begin{quote}\it
Extreme events are often quite different from ordinary everyday behavior and ordinary behavior often has little to say about extremes: Then only extreme events give useful information about future extreme events.

Theoretically motivated statistical models give much better possibilities to learn from experience (and compare) than if everyone uses her/his own ad hoc method.
\end{quote}
\vspace{-2ex}
This quote illustrates well the philosophy of statisticians working on extremes. Extreme Value Theory (EVT) provides powerful asymptotic theorems, as useful for describing tails behaviors as is the Central Limit Theorem for the mean behavior. Having general approximative distribution(s) describing extremes allows for better understanding and predicting extreme cyber risk, extrapolating from data general behaviors. 
Analysing extremes in cyber risk data has been an important focus of several authors; results will be discussed in this review, when looking at various models types, in Section~\ref{ss:models}. Before, let us  give in the next section a brief synopsis on the research pillars for studying cyber risk. 

\subsection{Research on cyber risk}
\label{ss:research}
\vspace{-2ex}

The exponential growth of information technologies and cyber attacks in parallel, have brought cyber research on the front line, mainly on cyber security due to governmental stakes and investments, but also on cyber risk. Due to its various specificities, it represents a big challenge for researchers in hard sciences and humanities, who have to investigate the topic in their own discipline but within a multidimensional framework. 
It was our goal when co-editing a special issue on the topic (for another journal), inviting researchers from different fields to contribute with their own expertise, e.g. from forensic science (\cite{Barlatier2020}) or with a graph mining method for approaching insurance ratemaking (\cite{Yeftanus2021}), among others (see \cite{Dacorogna2022a} for a presentation of the papers enclosed in this special issue).
Risk management is by nature pluridisciplinary. While quantitative risk management relies mainly on statistical modelling, due to the random nature of risk, the difficulty will be to include other considerations in the models. The latter  are the motivations of the wide reaching of this paper. 

In the last few years, the literature on the subject has become so vast in many research fields, that providing an exhaustive critical review (even with AI tools) is rather difficult in a reasonable amount of time. Our focus in this section is on cyber risk. As there are already a few good surveys on it concentrating on business and actuarial sciences \citep{Eling2020}, or on actuarial models and pricing \citep{Awiszus2022}, we will select results about some issues that we find important so far, refering also to the state of the art given in \cite{Dacorogna2022}, pointing out some research directions and their limitations. Complementary and more exhaustive information can be found in the cited surveys and in other papers.     

\paragraph{Taxonomy of cyber risk}

Different taxonomies are built according to the expertise field and to which purpose it is aimed at. For the governmental entities, they need to be precise and detailed, to better understand the context. For instance, ministeries of justice have generally a granular classification of cybercrimes (e.g. 475 types in France\footnote{See \url{www.justice.gouv.fr/include_htm/pub/rap_cybercriminalite_annexes.pdf}}) for a fair understanding and judgment. In forensics, \cite{Phillips2022} came up with a classification of cyber crimes as a tree, with a root of three categories (in fact, corresponding to the three layers of cyber space), accepted by the European Commission in 2013 with a slightly different terminology, namely:
\vspace{-2.5ex}
\begin{enumerate}
  \item Crimes against machine (Offences unique to computers and information systems, for EU)
  \vspace{-1.5ex}
  \item Crimes using the machine (Traditional offences, for EU)
  \vspace{-1.5ex} 
  \item Crimes in the machine (Content related offences, for EU)
\end{enumerate}
\vspace{-2ex}  
The police and gendarmerie, involved in the prevention of individuals and companies against attackers, are using the justice ministery classification, but also built their own, regrouping some crimes in one class, according to their understanding and specificity. 
Turning to another type of protection, the insurance, classification of course cannot be as granular, and is based on types of exposures, on which insurers study the statistical properties. 
Statisticians may look into data the other way round, regrouping observations having similar patterns, parameters or statistical characteristics, to deduce some classification. It is what is done e.g. in \cite{Farkas2021}, using regression trees based on CART (Clustering And Regression Trees) algorithm rather for the central part of the severity distribution, then for the tail part. In \cite{Dacorogna2022}, it is suggested to base a classification on the fatness of the tail of the severity distribution, studying each type of cyber crimes and collecting those presenting a similar fatness.    
\cite{Agrafiotis2018} provide a taxonomy of cyber harms regrouping them in five broad categories: physical or digital harms, economic harms, psychological harms, reputational harms and social and societal harms. Then,  they study the possible consequences of the various categories. Their approach is rather descriptive but the authors offer a good list of possible harms. Experts working for the big insurance broker AON, \cite{Cohen2019} suggest a taxonomy parallel to operational risk. Looking at two sets of data private to AON and public, they argue that cyber risk is an operational risk. \cite{Eling2019} adopt also a taxonomy based on operational risk in their study. Towards insurance, the most important effort to classify cyber risks based on exposure is due to the~\cite{CRO2016}. The latter is now prevailing in the insurance industry. Note also that those taxonomies are not fully compatible; some more work is needed. 

\paragraph{Debate on the type of risk}

Only a few datasets on cyber attacks or crimes come from open sources, namely the public Privacy Rights Clearinghouse (PRC) database, 
or, more recently, the database constituted by \cite{Eling2019}, extracting 1,579 cyber risk incidents from an operational database. 
The free access to the PRC database explains its use in most academic studies on the topic, as in the papers already mentioned, or in \cite{Bessy-Roland2021,Eling2017,Jamilov2021,Poyraz2020,Wheatley2021}, to list a few more. Nevertheless, most of those studies (with the exception of those by Jamilov et al. and Wheatley et al.) are plagged by the difficulty to attach an amount of money (not given in the database) to the number of data breaches.
Other used sources are non-public, as the database compiled by AON, containing 30'000 cases, used by \cite{Cohen2019}, or databases built and owned by companies to develop commercial models, or databases as that of the French Gendarmerie Nationale, or of \cite{Lucy2022}, made available to academics via partnerships.

Various statistical methods are either applied or developed on those databases. They do not always provide converging results, which makes it difficult to find a concise and confident view on the characteristics of cyber risk. For instance, \cite{Eling2019} conclude that cyber risks differ from other operational risk categories, contrary to~\cite{Cohen2019}. This divergence may be due to various reasons, among which might be the poor quality of the underlying database being used as representative of the risk. 
It highlights how much data are crucial and needed. But it is certainly difficult to judge about the representativeness of databases, so, one may circumvent this issue and try to analyse as many datasets as possible, for a larger comparison. If not enough datasets are made available, one may try to build new ones (as \cite{Eling2019} did), using for instance medias reports, or finding other ideas. 
\begin{table}[h]
  \begin{center}
  \caption{\label{tab:tailIndex} \sf\small Comparison between estimated tail indices, presented by decreasing sample size (except for the last one). The sources are \cite{Dacorogna2022} (DDK 2022), \cite{Eling2019} (EW 2019), \cite{Cohen2019} (Aon 2019), and \cite{Farkas2021} (FLT 2021), respectively.}
  \vspace{0.5ex}
  \begin{tabular}{l@{\hskip 30pt}c@{\hskip 30pt}c@{\hskip 30pt}c@{\hskip 30pt}c@{\hskip 5pt}}
      \hline
      &&&&\\[-1.5ex]
  Paper     & DDK 2022 & EW 2019 &  Aon 2019 & FLT 2021 \\[0.5ex]
  \hline\hline
  &&&&\\[-0.5ex]
  Data Source & French GN & SAS OpRisk$^*$ & Aon Force$^+$ & PRC$^{**}$  \\
  Number of Observations  & 60,985 & 1,579 & 376 &  6,160  \\
  Loss amounts  & $\geq 0$ \EUR &  $\geq 100,000$ US\$ & $\geq 1$ mUS\$& $\dagger$ \\
  {\it Tail Index $\xi$}  & {\it 0.81} & {\it 0.96} & {\it 0.50} & {\it 2.16} \\
  {\it Shape parameter $\alpha=1/\xi$} & {\it 1.24} & {\it 1.04} & {\it 2.00} & {\it 0.46}  \\
  95\% Confidence Range (CR) for $\alpha$ & [1.21 ; 1.26] & [0.84 ; 1.36] & NA & [0.42 ; 0.51] \\[1.5ex]
  \hline
  \end{tabular}
  \end{center}
  \vspace{-1.5ex}
  \sf\footnotesize
  *) SAS OpRisk Global data\\
	**) Privacy Rights Clearinghouse\\
  +) Publicly Known Matters/AON Claims\\
  $\dagger$) loss extrapolated from the number of records of data breaches
  \end{table}

First statistical results are given to answer a crucial question asked by insurers, namely is cyber risk insurable? A necessary condition for it is to have a finite expectation for the loss. This can be judged from the fatness of the tail distribution, measured by the tail index (or its inverse, the shape parameter). Recall, from EVT, that distributions can be classified into three types, according to the asymptotic behavior of their tail distribution: a tail index equal to 0 (exponential decay of the tail distribution) corresponds to the Gumbel maximum domain of attraction (MDA), a positive tail index (power decay) to the Fr\'echet MDA, and a negative one to the Weibull MDA (see any textbook on EVT, or \cite{Kratz2019} for an introduction to the topic). The fatter the tail, the less existing moments. For instance, if the shape parameter $\alpha=1/\xi$ is between 0 and 1, it means that all moments (in particular the expectation) are infinite; if $1<\alpha \leq 2$, the expectation exists but not the variance, etc. In Table~\ref{tab:tailIndex}, we display, for comparison, recent results obtained on the estimation of the tail index for four different databases. We see that three out of four come up with a tail index $\xi$ less than 1 (or shape parameter $\alpha=1/\xi$  above 1) meaning that the expectation exists ({\it i.e.} is finite). There is only one, \citep{Farkas2021}, that results surprisingly in a tail index much larger than 1 ({\it i.e.} infinite expectation). However, the study is not performed directly on loss amounts, but on the number of records of data breaches transformed into loss amounts via an extrapolation of a model by \cite{Jacobs2014}.
In his communication, Jacobs suggested a log-log regression for relating number of breaches to loss amounts on 2014 data gathered by the Ponemon Institute LLC, improving the method used by Ponemon. Nevertheless, the model is far from being satisfactory, as commented by Jacobs: "[...] it is painfully clear that there are a lot more factors contributing to loss than just a count of records lost." 
This may explain the gap observed in the last column of Table~\ref{tab:tailIndex}, with a tail index estimate of 2.16 (!) when fitting a Generalized Pareto Distribution (GPD)  on the whole set of observations, or between 1.43 and 3.26 when fitting the GPD on leaves. The authors themselves criticize their results: "This indicates that the quadratic based regression method may not only lack robustness, but lead to ill-defined estimates (since the conditional expectation is not defined, at least for some leaves in the tree)". Here, more investigation would be needed to see how to use the information given by the number of records of data breaches to properly value this typical intangible. Therefore, we discard this case to conclude from Table~\ref{tab:tailIndex} to a finite expectation of the loss amounts. Further studies (among which, the one we are currently conducting on ransomwares) will help consolidate estimates of the fatness of the tail distribution.  

\subsection{Types of cyber risk models}\label{ss:models}
\vspace{-2ex}
We divide models for cyber risk into five types. This categorization, made in terms of methodology rather than risk properties, reflects our point of view on the topic. It does not mean that models could not have been built combining those methods. For instance, we find it really important not to oppose Artificial Intelligence techniques (part of statistics and based on probability and statistics theory) with probabilistic modelling. Combining both may help develop very relevant models, which can be dynamically calibrated. Our five types are: actuarial models, stochastic models for risk contagion, data driven models, exposure models and game theory based models. Other legitimate ways of classifying the models exist in the literature, e.g. \cite{Eling2020,Awiszus2022}, with some classes common to ours, which we discuss only when having additional points. 

We perceive two main streams in insurance companies: (i) one relying on commercial models, (ii) the other on actuarial models based on open academic research and data they collected themselves.
 \\[-5ex]
\begin{enumerate}
  \item {\it Actuarial models} (severity-frequency), applied to loss data; they are often named by actuaries {\it experience models}.
This class being common to the recent comprehensive surveys by \cite{Awiszus2022} and \cite{Eling2020}, we refer to them as their presentation corresponds to our views. Moreover, the models are described mathematically in details in \cite{Awiszus2022}, which provides a nice complement. Let us also refer to the discussion given in \cite{Dacorogna2022}[Section 2 - state of the art] on the following papers: \cite{Carfora2019,Romanosky2019,Zeller2020,Bouveret2018}.
%
  \item {\it Stochastic models for risk contagion.}
  One of the characteristics of cyber risk is that it takes place in a wide network of computers and links, which makes it susceptible to be modelled using network (or graph) models, epidemiologic/pandemic models, or other adequate stochastic models. This class corresponding exactly to that of  \cite{Awiszus2022}, we simply refer to their Sections 3.1 and 3.2. There, the models are nicely presented. A discussion follows on the various papers applying some of those models, which are calibrated to cyber risk (on real or simulated data). We invite the reader to look at these references as we do not list them here, but they constitute an important step in the understanding of cyber risk literature.
  \item {\it Data driven (AI) Models.}
  This class may overlap the next one of exposure models, but we refer to examples of academic papers where the methods and data are clearly stated, contrarily to commercial models using AI techniques. In such academic papers, the innovation may be either on the choice of input data to study a specific subject, or in the development of AI methods.  We give an example for each case. Note that, as already mentioned, AI techniques can be combined with classical models to improve the prediction of the models. 
  
  In \cite{Shu2018}, the authors explore social media as a means to understand and predict cyber attacks. They relate the latter to social behaviors and perform a sentiment analysis using emotional signals common in social media, such as emoticons or punctuation. For that, they develop an unsupervised algorithm for sentiment prediction, without requiring labeled sentiment data beforehand, and taking into account emotion words and emoticons, as well as their correlations. They introduce the sentiment predictors in a logistic regression for evaluating probabilities of attacks.
The effectiveness of their resulting model is tested on real-world Tweet data related to several cyber attacks by a well-known hacker group.

While this first example proposes a new methodology to predict probabilities of cyber attacks, the second example, \cite{Subroto2019}, is an application of AI techniques (statistical machine learning and artificial neural network) to build a confusion matrix on big data to predict cyber attacks, but with a relative short time lag between prediction and occurence of attacks (so, difficult to exploit). 

  \item {\it Exposure models}. Those models are based on exposure to cyber risk and stress scenarios on this exposure. Exposure base is difficult to determine as software and systems are complex and highly technical, and the exposure changes rapidly. 
  
Those sophisticated models are mainly commercial ones. Those firms have been either specialized in NatCat models, as e.g. {\it RMS}, or {\it Verisk} (former AIR), or companies specialized in software for insurance like {\it Guidewire} (including {\it Cyence}), {\it Cybercube}, or {\it DeNexus}, or some combining modelling, security, and insurance, as {\it Coalition}. 
  
There are three main approaches used by those firms to gather exposure data: (i) mapping the clients network; (ii) grasping cyber data from internet; (iii) building honeypots to lure hackers and extract information on their strategies and on the possible weak spots of the clients system. Of course, a combination of these three sources is also possible and used by some, as e.g. Coalition. Based on this exposure, scenario simulations are developed to evaluate probability distributions of system failures and their severity.

Through what we could hear in conferences or workshops on the topic, commercial models seem to be ahead compared to academic research, in terms of innovative methodologies. Their heavy use of big data collected over internet and cyber security techniques is a big advantage, as well as their attractiveness for young PhDs due to the good research environment they offer. However, the methods they develop are not of easy access: Due to the strong competition between those companies, the level of transparency is pretty low, which may hinder scientific progress. In particular, we lack statistical evidence of the effectiveness of those models. 
Nevertheless, some very innovative companies have decided to play the game of scientific openness and publish their scientific research.  For example, reinsurers like Swiss Re or SCOR publish many scientific papers. Moreover, regulators require companies to understand and explain in detail their models in view of their approval. This certainly pushes commercial companies to be more transparent towards their customers, thus also towards the scientific community. Finally, the relation between academics and professionals, thanks to the advent of risk based regulation, also increased, which is key for understanding this complex risk as pointed out in (\cite{Dacorogna2015}).  
  \item {\it Game theory based models.} 
  
Cyber risk is fundamentally different than natural catastrophes: It is man-made, so, studying the motivations of the attackers could be a way to look at the problem using game theory. The latter actors are not directly useful for pricing insurance, which may explain why hackers have not been taken into account in \cite{Marotta2017}, or \cite{Awiszus2022} where this class of models is also suggested. Looking at hackers' motivation may be a useful study for governmental or legal entities too, to draw strategies and priorities to protect individuals or organizations.     

Including hackers as a fourth actor in game theory based models should bring valuable insights in risk management. It helps identify within an organization the most important targets and analyze attackers motivations. Indeed, it would allow concentration on the protection of the most targeted assets, while hackers motivations help detect changes in the political environment that may enhance the risk of attacks. This, in turn, will be addressed by modifying the probability distributions in certain situations, and thus the insurance pricing. 
 
Note that studying hackers' motivations is often done via data driven models (based on machine learning, or more generally AI-tools). Here also, we see that combining data driven/AI and other types of models may improve our understanding of the risk. For instance, the building of honeypots allows to gather data that could be used to fit the game theory based model and help prioritize threats and, as said by T. Henriques and S. Bell from Coalition: "learn about cyber criminals behavior without paying claims". Cyence (Guidewire) also uses technical and behavorial data available about cyber threats to modulate their cyber risk probability distribution, using AI tools. Another example can be found in \cite{Yue2019}. They study the impact of online-hacker-forum discussion on the extent of distributed denial of service (DDOS) attacks. Their main result is that, following such forums helps offset the harm caused by malicious sharing. The authors argue that paying closer attention to the flow of public discussions instead of focusing on disclosure of malicious information per se, may be a rich source for both firms and regulators.

So far, game theory is mainly used in cyber security and in commercial models. In \cite{Musman2018} Cyber Security Game takes into account the widespread interconnectedness of cyber systems, where defenders must defend all multi-step attack paths and an attacker only needs one to succeed. It employs a game theoretic solution using a game formulation that identifies defense strategies to minimize the maximum cyber risk. This paper discusses the methods and models that compose Cyber Security Game. Another example is \cite{He2020}. The authors apply discrete game-theoretic analysis to the defense of correlated physical systems. In the case of cloud computing infrastructure, they show that, as one would expect, the defense is higher if the attackers have no information about the distribution of the servers. Those are examples of approaches that help better design resilient systems.
\end{enumerate}

Often, people object that modelling is difficult, even impossible, because cyber landscape is changing rapidly. However, as already commented in the introduction, the scientific approach is to detect invariance in phenomenon; it does not mean using only a static point of view. In fact, a given dynamics can correspond to an invariance. Our brief survey shows that progress in finding those invariances is under way, even though there is still a lot to be done. As actuaries, our task is to model the risk. Moreover, data driven / AI models and probabilistic models must not be opposed but combined and should learn from each other. It is fundamental not to lose sight of the understanding of a process, the primary goal of science, and not only to aim for prediction, the usual target of AI models. Particularly in a fast changing environment, data driven approaches are plagued by the non-representativeness of some data. Thus, a deeper understanding of the underlying process helps selecting the right data for the AI tools.

Many models for cyber risk already exist, often borrowed from existing literature of probabilistic models and adapted to the specificities of cyber risk. This is a necessary first step in the advancement of research. So, at the moment, there is nothing special in cyber risk models and we know their limitations. This is why we need to look for more data, whatever the way, and study on data the empirical regularities of this risk to be able to judge if those models are reasonable or if more innovative approaches to modelling are required to master such a risk. It is also the reason why we address in this survey this risk in multiple ways, with a wider risk management point of view rather than a pure actuarial one. We hope it will stimulate innovation in modelling.

\section{What is so special about cyber risk management compared to other types of risk management?}
\label{sec-riskmgmt}
\vspace{-2ex}
In the previous section, we reviewed various modelling approaches to quantitatively assess cyber risk. In this section, we widen our purpose to risk management, both quantitative and qualitative, to explore other ways of studying this risk. This will allow us to propose new avenues for developing our quantitative approach to better understand cyber risk in the future. 

\subsection{A big risk to manage}\label{ss:big-risk}

Empirical studies on data confirm the common fear that cyber risk belongs to the class of big risks, as natural catastrophes (often named NatCat), with a strong systemic component. However, it differs from NatCat as it originates directly from human actions (using the insurance terminology, it is not "an act of God"). It requires management rules similar to those of NatCat, but must combine various aspects depending on the three layers defining the cyber space (see Section~\ref{ss-space}). This induces the development of a variety of rules adapted to each layer, but also articulated between each other through those layers. For instance, on the psycho-cognitive level, insurance companies are working to define the notion of cyber war to be able to reformulate war exclusions fitting the dematerialized nature of cyber. It also means that, when excluding cyber war, they will consider insuring for instance ransomware that affect the two other layers, hardware and software ones.

A second example is governments developing special agencies to tackle the problem of cyber harassment, adapting the legal procedures for fighting it, and also providing software tools (for instance to educate people on cybersecurity and make them aware about means to fight against cyber harassment). Examples of these agencies and information sharing platforms are ANSI in France, MELANI in Switzerland, BSI in Germany and ENISA at the European level. Another example coming from the management of NatCat or terrorist attacks would be to develop public funds that cover the systemic aspect of cyber risk, to which every private insurance company should contribute. This is also a recent proposal put forward by the CEO of {\it Zurich Insurance}, one of the largest insurance companies in the world.  

Coming back to the qualifier "big risk", it is to be understood in terms of major impact (e.g. in terms of risk capital, loss, reputation, psychology, ...). For instance, considering the financial impact to cover such a risk, we display in Table~\ref{tab:risk-measures} the ratio of Expected Shortfall (ES), over the mean, using several methods to estimate the tail of the  underlying loss-distribution (see~\cite{Dacorogna2022}). This ratio is related to the capital intensity (=capital/premium), usual measure in insurance for classifying risks from attritional to large, then to catastrophic ones.
Recall that a simplified expression for the technical premium $P$ corresponds to $\displaystyle P=\mathbf{E}[L](1+e+\eta\,f)$  where $\mathbf{E}[L]$ is the expected loss (estimated by the empirical mean), $e$  the expense ratio, $\eta$ the cost of capital, and $f=\rho(l)/\mathbf{E}[L]$ with $\rho(L)$ the risk measure for the loss. Thus, the capital intensity, denoted by I, satisfies $I=f/(1+e + \eta\,f )$, or, equivalently, $f=\frac{(1+e)I}{1-\eta I}$ (with $I<1/\eta$). 
In Table~\ref{tab:risk-measures}, we display the factor $f$. For instance, for $f=20$, with $e=35\%$ (for insurance including broker fees) and $\eta=15\%$, then $I=4.6$. This is of course the capital intensity for a standalone risk, {\it i.e.} without considering diversification of the insurance portfolio. When considering the latter, the capital intensity will drop considerably. Note that a capital intensity bigger than 1 is considered by insurers as a catastrophic risk.  
\begin{table}
    \centering
    \caption{\label{tab:risk-measures}\sf\small [Table 11 in \cite{Dacorogna2022}]. Estimates $\widehat{ES}(p)$ of Expected Shortfall $ES(p)$ (as computed in Equation (11) in the quoted paper) for $p=97.5\%$ and $99.77\%$, expressed as the multiplying factor of the estimated mean (which value is 3476~\EUR) for various models. Comparison with the empirical values $\widetilde{ES}(p)$ (also expressed as the factor, which multiplied by the mean gives the evaluated risk measures) by computing the relative variation $\Delta$ in $\%$. \vspace{0.7ex}}
  \small
  \hspace*{-20pt}
      \begin{tabular}{l@{\hskip 60pt} c c@{\hskip 50pt} c c}
      \hline
          &&&&\\[-1.5ex]
          Factor $f$ for risk measures:  & $\widehat{ES}(p)$ &                  & $\widehat{ES}(p)$  & ~~~ \\
                                         &     $p=97.5$\%    & $\Delta$ (in \%) &     $p=99.77$\%    & $\Delta$ (in \%)\\[0.5ex]
      \hline
      \hline
          &&&&\\[-1.5ex]
    {\it Empirical}   $\widetilde{ES}(p)$  &     {\it 23}    &                  &  {\it 114}         & ~~ \\[0.5ex]
  \hline
	
          &&&&\\[-1.5ex]
  \footnotesize{\cite{Dacorogna2022} ($\alpha=1.24$)} &       19        &       -17.1      &       132          & ~15.9 \\
  \footnotesize{AMSE ($\alpha=1.17$)}    	&        43        &       ~85.1      &       331          & 190.6 \\
	\footnotesize{Danielsson-al.(01) ($\alpha=1.15$)} &  47    &       101.7      &       373          & 227.1 \\
  \footnotesize{Hall (1990)($u_2\!\!=\!\!q(99.45\%)$;$\alpha\!=\!1.37$)} &   8 & 21.4 & 159 &  39.9 \\
   \footnotesize{Hall (1990) ($\alpha=1.61$)}  &   $-$   & $-$ &   119  &  ~~4.2  \\
  \footnotesize{Reiss \&Thomas(07) ($\alpha=1.47$)}& $-$ & $-$ &   130  &  ~14.2 \\
    \hline
      \end{tabular}
  \end{table}%

(Re)insurance industry has already gained much experience in managing extreme risks for NatCat and recently for pandemics; this should help for managing cyber. Two important lessons learned so far are strive for diversification in the portfolio, and join forces between private and public for tail risks, since extreme risks are difficult (if not impossible) to diversify.  If geographical diversification is absent in the cyber case, some other ways of diversifying may be possible, e.g. operating systems, redundant networks. Moreover, as already pointed out, insurance companies should incentivize their customers to invest in cyber security to improve their resilience to cyber attacks. 

\subsection{From cyber security to cyber resilience}
\label{ss-resilience}

The notion of cyber resilience has lately gained momentum due to the repetitive failures of cyber security systems to prevent attacks. This concept recognizes that breaches will always occur, thus organizations need to concentrate their efforts on surviving those breaches, while improving their ability to detect attacks. Resilience may be defined as "`the ability to continuously deliver the intended outcome despite adverse cyber events" (\cite{Bjoerck2015}) or as "The ability to anticipate, withstand, recover from, and adapt to adverse conditions, stresses, attacks, or compromises on systems that use or are enabled by cyber resources." by the US National Institute of Standard and Technology (NIST). This means for organizations to "change their security posture from a defensive stance focused on malware to a more realistic and resilient approach" as advocated by Symantec, a company specialized in cyber security products (\cite{Symantec2019}).

Cyber resilience implies designing a cyber strategy based on the current (and expected future) threat environment, taking into account processes and technology, but also all actors, people (individual and professional) and organizations (private and public), according to the acceptable risk level for each. It goes from education to dedicated processes, going through implementation of performance measurements, even at a governmental level. 

While we are going to derive this concept with an actuarial point of view, it is worth recalling how combining the expertise from all fields is needed. All governmental institutions are clearly aware of that, and take part in building a more cyber resilient environment. Let us take the example of the European Union, preparing a legislation on cyber resilience, named the `cyber-resilience act proposal' (see \cite{EU2022}). Quoting U. von der Leyen: {\it If everything is connected, everything can be hacked. Given that resources are scarce, we have to bundle our forces $\ldots$ This is why we need a European cyber defence policy, including legislation setting common standards under a new European cyber resilience act.} The same concern holds worldwide, e.g. the \cite{Wef2022} wrote a white paper on cyber resilience, stating again that this topic should involve the entire society. They also came up with the proposal of a Cyber Resilience Index (CRI) using a weighted scoring of 64 performance measures (on all aspects) collected for each organization with a self assessment (see Figure 4 in the cited document).

What does cyber resilience include? Five important priorities and assessments emerge from the point of view of an organization, namely:
\vspace{-2.5ex}
\begin{itemize}
  \item Becoming better at stopping the attacks: It means to invest in prevention through cyber security measures, education, awareness, $\ldots$ 
  \vspace{-1ex}
  \item Detecting faster breaches (detection). Currently, it takes much too long to detect a breach: The average time to identify a breach in 2022 was 207 days according to a study conducted by \cite{IBM2022} over 550 organizations.
  \vspace{-1ex}
  \item Fixing faster breaches through agile reactions and good preparedness. Here also, improvement is desirable: In the same IBM report, the average lifecycle of a breach is reported to be of 70 days from identification to containment.
  \vspace{-1ex}
  \item Reducing breach impact by improving the system. The average cost of a data breach is $4.35$ million USD as of 2022 (compared with $3.86$ million USD in 2020); \cite{IBM2022}. 
  \vspace{-1ex}
  \item Ensuring company's access to cash, to help their resilience. For instance, ENISA interviewed European SMEs during the pandemic\footnote{See \url{www.enisa.europa.eu/topics/cybersecurity-education/sme_cybersecurity}}; among those, 90\% stated that cybersecurity issues would have serious negative impacts on their business within a week of the issues happening, with 57\% saying they would most likely become bankrupt or go out of business. In general, this is due to a lack of cash to pay bills and employees. An insurance covering this type of event would definitively increase resilience; it is discussed in the insurance section. 
\end{itemize}

\subsection{Various strategies}

In the previous sections, we identified the need to manage cyber risk similarly as a catastrophic risk and to build cyber resilience beyond cyber security. In this part, we consider various components that help build such a resilience. 

\paragraph{Causes versus consequences}

As seen previously, modelling can be tackled when looking at consequences only, and not necessarily caring for the causes, which are more difficult to apprehend. When turning to risk management, both sides are needed. Identifying the possible targets of cyber attacks, e.g. secured and confidential data or intelligent energy management systems, may help getting better prepared to fight against malicious attackers.

In Section~\ref{sec-modelling}, we discussed the problem of classifying  cyber crimes in view of risk analysis. We need to complement this discussion questioning the various motivations of cyber attackers, as well as the types of cyber attacks. Indeed, a taxonomy of the various motivations of cyber attacks may also point out the weak spots of a system and lead to design specific strategies to prevent or limit such attacks. Those motivations, part of the third layer (psycho-cognitive) of the cyber space, will use tools of the first two layers and impact the three layers. 

In \cite{Cybercube2022}, the authors analyze the various types of attackers and their motivations, to better understand them. It should help fight against cyber crimes, according to the authors, who say: "A greater understanding of the key cyber actors will help the insurance sector predict how and where future attacks could arise and inform estimations of attack frequency and severity".
Three categories of attackers are suggested: state-sponsored actors, organised criminal gangs, and hacktivists seeking to drive social change, and their privileged tactics depicted. In a short paper, \cite{Han2014} give a similar classification as the one in \cite{Cybercube2022}, but make also an important distinction between insider and outsider attackers. 
Another threat, analysed in \cite{Guembe2022}, is the increased use of AI techniques to drive attacks, which might render inadequate the existing cyber defense infrastructures. In such a case, analyzing causes of such attacks will become very difficult; one would have to rather study their consequences. \cite{Li2017} makes a comprehensive list of 29 motivations of illegal activities in the cyberspace. Besides the common motivations found in the literature, as e.g. `Hacking for acquiring financial gains' or `Sexually motivated misuse', more unusual ones appear in the author's list as `Practising and show up programming skills' or `Realizing free expression of ego'. 

\paragraph{Cyber security versus insurance}

Risk is the common thread between cyber security and insurance. While the first aims at reducing the risk of cyber attacks and protect against the unauthorised exploitation of systems, networks, and technologies, the second hedges the residual risk. Hence, insurance is an important component of building up resilience, together with education, recovery processes, and legal enforcement.

This is why it is crucial to draw the risk profile of any organization, to determine the amount of investments in building its resilience. It means evaluating both qualitatively and quantitatively the current risk and decide what is the acceptable amount of risk to be carried internally and what should be passed externally. Then, the question for management arises, where and how to allocate resources between cyber security and cyber insurance. It is what~\cite{Marotta2017,Wang2019,Hoffmann2020} explore in their papers. However, their models are not supported by strong empirical evidences due to the lack of data of various sources, but research on this subject will go on. Another interesting perspective for the use of insurance in conjunction with cybersecurity is given in~\cite{Shetty2018}. There, they highlight the insurer's incentives for insured companies to invest in the self-protection of their IT infrastructure as an important benefit for risk management.
  
Currently, management relies much more on security spending rather than on insurance covers. Investment in cyber security is expected to reach (worldwide) 124 billion US\$ in 2022 compared to insurance premiums of 8 billion US\$; see \cite{Accenture2019}. One of the reasons for this discrepancy is given in the same report: "There is a lack of capacity in the market and a willingness of insurance companies to take over this risk". However, another reason can be found in a \cite{Wef2022a} survey, where it is stated that 59\% of cyber leaders do not distinguish between cyber security and cyber resilience. It would imply that, for them, insurance covers are not clearly part of cyber resilience, pointing out an opportunity for insurance companies to better communicate and play an active role here. Indeed, insurance companies do provide cyber cover but are limited in their coverage and reimbursements. It may be explained by the fact that cyber risk is not yet fully understood and that the fear of systemic risk among insurers is very high (see Section~\ref{ss:big-risk}). We will come back on this issue in the next section.

A big incentive for investing in cyber security and research comes from the new regulations that governments are putting in place, in many places, particularly in EU (with the General Data Protection Regulation - GDPR) and USA (with state regulations as e.g. the California Consumer Privacy Act). It also raises demands for insurance covers. Research on cyber security (for a comprehensive review of the standard advances in this field, see for instance \cite{Li2021}) is going along with research on cyber risk (see Section~\ref{sec-modelling}). A good understanding of each will be useful for coming up with management guidance on the multidimensional problem of resilience. 
 
Insurance companies have played a big role in the past in terms of prevention, for instance of fire risk, but also in setting up traffic rules when developing motor insurance. Here as well, they could play a similar role in terms of prevention, working with cyber security firms without replacing them as advocated by some actors. However, some new entrants, as for instance {\it Coalition}, are combining cyber security services with insurance backing. They advertize the rapid feedback they can give to their customers about the possible threats and protection against them, while making their most responsive customers benefit from reduced insurance premium. Yet, the risk of combining cyber security and insurance in a same company is the accumulation of one risk in insurance terms, thus losing diversification benefits.  

\paragraph{Development of cyber insurance}

Munich Re, the world's largest reinsurance company and also the largest cyber risk reinsurer, recalls\footnote{\url{www.munichre.com/topics-online/en/digitalisation/cyber/cyber-insurance-risks-and-trends-2022.html}} that the demand for cyber insurance has grown greatly in recent years, and sees cyber premiums worldwide standing at 9.2bn US\$ in 2022 (while at 4.7bn US\$ in 2018) and estimates that they will reach a value of approximately 22bn US\$ by 2025. The share of premiums is geographically very unbalanced, half of it originating from USA. Cyber insurance market is growing but remains very small in comparison with total insurance premiums, estimated by Allianz to be at 4,000bn US\$. It reflects the reluctance of insurance companies to fully enter in this market, despite the strong demand (in the same quoted article, MunichRe reports that 83\% of their interviewees report that their company is not adequately protected against cyber threats). Generally, individuals, companies or organizations are offered inadequate insurance covers against cyber risk due to lack of knowledge by the companies and their fear of a strong systemic component (see e.g. ~\cite{Advisen2018}~or~\cite{SwissRE2017}).

Another reason for the weak offer by insurers is the lack of - or poor - cyber security of SME's or individuals. Allianz announced that they refused 3/4 of the demands for cyber covers because of lack of cyber security. This is again a sign that insurance companies play an important role in incentivizing SMEs to invest more in cyber security. They demand a high level of security before granting a cover. In a recent study by the French risk association AMRAE, they report 44\% volume increase in premium while underwriting capacity shrank for almost all companies (deductibles rose to EUR 4m for large companies). This was the consequence of a jump in claim amounts from EUR 73m in 2019 to EUR 217m in 2020, while the premiums were only growing from 87m EUR to 130m EUR. In 2021, the profitability returned, but only for large companies, with a loss ratio (loss over premiums) of 58\%, while of 261\% for medium and 325\% for small companies. Luckily for the French insurance industry, large companies represented 83\% of the premiums, resulting in an overall loss ratio of 89\%; see \cite{Lucy2022}.

Thus, cyber risk is an opportunity for growing the insurance market, particularly in developed countries where this market is otherwise saturated. It constitutes another important characteristics of this risk. However, some prerequisites for a sustainable cyber insurance market are:
\vspace{-2.5ex}
\begin{itemize}
  \item Companies enforcing underwriting discipline will ensure acceptable volatility and profit levels. The risk is that new entrants would drive soft coverage terms and low prices. It is not the case anymore as the costs are rapidly rising, as pointed out above.
  \vspace{-1ex}
  \item At policy holders level, if event frequencies are too high, cyber will become uninsurable, thus the need of introducing cyber security as a prequisite when underwriting cyber risk. There is here a convergence between insurers demands and governments regulation proposals. 
  \vspace{-1ex}  
  \item Within insurance companies, the ability to measure and manage risk accumulation is key. Indeed, unplanned accumulation could result in a major event 
  shocking the entire market.
  \vspace{-1ex}
  \item Removing the so-called {\it silent} exposures to loss or liability from a cyber-triggered event in other lines of business (e.g. Directors and Officers, Property Damages, Business Interruption) in favor of proposing specific cyber policies.
\end{itemize}
\vspace{-2ex}
Considering the advances in quantitative risk management, for transforming the uncertain outcome of cyber risk into a measurable risk, companies would need to:
\vspace{-2.5ex}
\begin{itemize}
  \item Collect and analyze data on cyber attacks
  \vspace{-1ex}
  \item Develop probabilistic models to assess the risk
  \vspace{-1ex}
  \item Partner with cyber-security firms to reduce adverse selection and with researchers to explore the domain of cyber risk. It is expected, over time, that the modelling of cyber risks will be able to mirror the sophistication of the modelling of NatCat risks
  \vspace{-1ex}
  \item Access commercial models, as done for NatCat. This has been instrumental for developing the insurance market for NatCat, making it possible to compare prices across various actors. It can as well stimulate academic research on the topic. Property catastrophe modelling took around 25 years to mature – it seems that it will be much shorter for cyber as more tools are available and experience gained on NatCat and Pandemics can be put in use here. For instance, companies like RMS, AIR (Verisk), are already building on that experience:  RMS launched its probabilistic cyber model in 2019, while a rival vendor AIR (Verisk) Worldwide released one in 2021 and large (re)insurers are taking the lead in this effort. 
\end{itemize}
\vspace{-2ex}
What are the main covers proposed so far by insurers? We propose a list classifying those covers by purpose, but some may be combined in specific insurance policies. For instance, a cover for "Breach costs" combines the items: Data and system recovery, IT-Forensic, and Crisis management.  
\vspace{-2.5ex}
\begin{itemize}
  \item Data and systems recovery: Pay cost of data and systems restoration, replacement if a hacker causes damage to websites, programs or electronic data
  \vspace{-1ex}
  \item Third party liability: Reimburse or fend-off third party claims, notifying customers/regulators
  \vspace{-1ex}
  \item IT-Forensics: Pay for forensics to assess and contain damage
  \vspace{-1ex}
  \item Data protection laws: Legal advice to act correctly after a breach
  \vspace{-1ex}
  \item Crisis management: Pay cost of crisis communications to contain reputational damage, credit monitoring of affected customers, psychological support for individuals in the case of private policies\\[-5ex]
  \item Ransomware: Pay the ransom
  \vspace{-1ex}
  \item Business interruption: It is usually in a single contract 
\end{itemize}
\vspace{-2ex}
What is the demand of the market? Complaints of customers on those covers are that there are too many exceptions in the contract and/or that claim settlement time is very long (this can be explained by the need for the claim manager to check the claims for appropriateness, particularly for intangibles). It can take 6 months to be reimbursed; but we have seen that, according to ENISA, 58\% of the SME's might be banckrupted after 1 week.
\begin{figure}[h]
  \centering
  \includegraphics[scale=0.6]{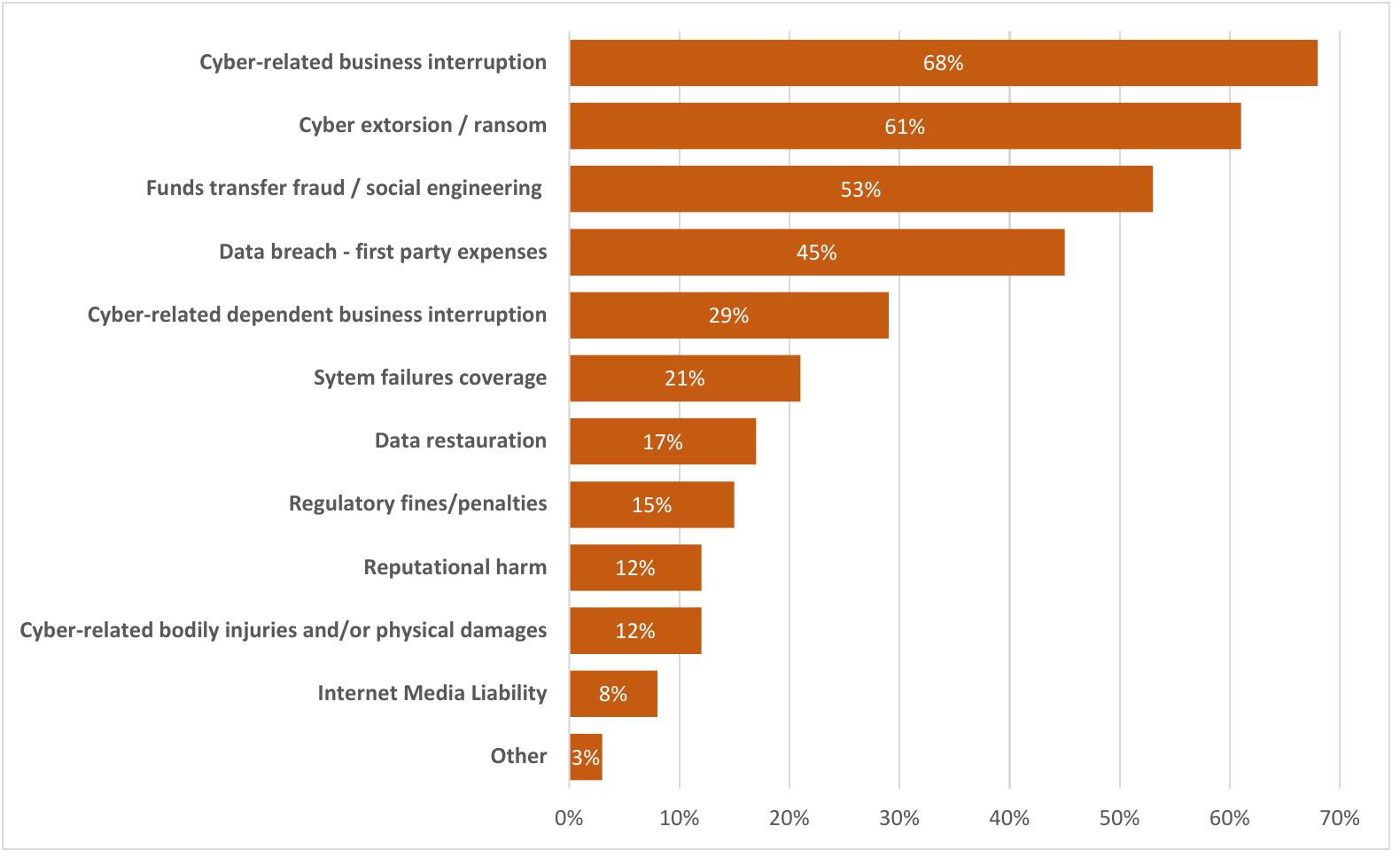}
  \parbox{300pt}{\caption{ \label{fig:Gallagher} \sf\small The results of a market survey on customer wishes for insurance covers done by Gallagher in February 2021.}}
\end{figure}
Many surveys of customer profiles show that the most important cover they need is against business interruption due to cyber attack. The results of Gallagher's survey in Figure~\ref{fig:Gallagher} show very well that more than 2/3 of the customers are worried about business interruption and dependent business interruption\footnote{(source \url{ https://www.ajg.com/us/news-and-insights/2021/jan/2021-cyber-insurance-market-report/})}. Ransom and financial extorsions are also big on customers' agenda, manifesting the increase in ransomware. 

After a cyber attack, the machines on the shop-floor stop and in the office nothing runs if the computers fail.
For the company to survive in such a case, the insurance should step in and pay an agreed daily rate until everything runs smoothly again, as advocated by some insurance consultants. This would compensate companies for the loss of profit, but, even more importantly, enable them to pay running costs. For avoiding bad risks, the insurer could couple his contract with a cyber security contract where the insurance would be triggered  after the intervention of the cyber security firm. Such an insurance cover is named {\it dependent business interruption}. Currently, it is usually not offered in cyber insurance. Besides answering directly the concern of SME's, we see three advantages for the insurer for offering such product to the insurance market: (i) It alleviates the problem of intangibles as the amount for the claim is fixed in the contract; (ii) The down-time of the system is confirmed by a third party (a cybersecurity firm); (iii) Taking the actuarial point of view, only consequences have to be considered whatever the cause and, moreover, data on business interruption are more abundant and of easier access than cyber insurance claims.

This is a good example of how to come up with an original solution based on traditional insurance covers like business interruption, adapted to this problem of cash needs.

  \paragraph{Risk transfer and pricing}

As for NatCat, one may transfer part of the extremes of cyber risk to investors. This was for instance suggested by Akinova or  Securis\footnote{Presentations made at the  ASTIN Cyber Working Group in May 2022}. 

For risk transfer, a consensus is clearly needed, as valuation models have to be agreed upon, by both parties. To look for such agreement, commercial cyber models are needed. So, we expect the development of those models to raise the interest of the Insurance Linked Securities (ILS) market for cyber, similarly as it was for NatCat. ILS have traditionally been seen as a diversifying allocation of investments. Until now, cyber losses have not been as correlated with other financial assets as feared. Hence, we should try to evaluate the amount of diversification such a risk offers to market with other types of risks. This is why the major challenge is to understand and model the dependence between cyber risk and other risks. 

We observe that the interest for cyber ILS is growing even though the market has not yet taken off. The question will be how to share the load between insurance companies, reinsurance companies, market, states, \dots Certainly, the experience gained in NatCat and also in life insurance will help designing a system to manage the need of capital for this risk.

\subsection{A political risk}

Due to its pervasive influence on society, there is an important component in cyber risk that is of political nature. Managing such a risk requires understanding the political drivers as they will bias the risk profile of an insurance portfolio. For instance, some industries could become more attractive targets for hackers; think about power plants in case of war (as seen in Ukraine), or hospitals during the pandemics. It may have consequences on both pricing of certain policies (modifying the distribution based on historical losses) and diversification of the portfolio (by modifying dependencies between losses among various industries). Due to its strong political dimension, media coverage plays an important role in cyberspace. In~\cite{Shandler2022}, the authors argue that the effect of cyber attack may have: "...insidious societal risks such as reduced trust in government", if the media coverage of the attacks are not well handled. Their conclusions come out of a thorough empirical study of survey data in the immediate aftermath of a ransomware attack against a D\"usseldorf hospital with 707 observations. Their conclusion is in line with our views that the political component plays an essential role in the context of cyber attacks. They write: `Cyber-attacks occur in a fundamentally different domain with unique qualities that alter traditional political outcomes'. They acknowledge the psycho-cognitive dimension (3rd layer) of the cyber space.

In a review article, \cite{Li2021} examine the current advances in the field of cyber security. They also warn that the cyberspace has created a phenomenon of power dissipation because of its characteristics of `low entry prices, anonymity, vulnerability, and asymmetry'. This means that governments, although playing an important role, are not the only actors in the power game of the cyber space. Other actors, such as private companies, organized terrorist and criminal groups, and individuals are now partaking in the game. National security has also widened its field of application: It is not only defined in terms of military issues and internal and external borders, but the declining of security on the cyberspace is a threat to national security and to the quality of life of citizens. This has another unintended consequence for the various theoretical approaches in international relations, based primarily on governmental actions. So, new ways of dealing with security are needed in the information age. The dependence on IT systems has become so pervasive all over the world for communication and control over the physical world that it is unimaginable that, for security reasons, some will separate from Internet. A half measure seems to be the development of sub-networks with limited access to the full Internet, as in China, Iran or Russia. At the same time, the effective control over the operations in the cyberspace is done by a relatively small number of individuals. Most of the users do not have the ability to modify or control the software or hardware they use. Paradoxically, even if only few people control the cyberspace, its decentralized nature currently hampers the technical ability to assign activities to individuals or groups or organizations, with a high degree of confidence. All these features make cyber risk management {\it special} and constitute the framework within which risk management must develop to ensure cyber resilience. Such management is time-consuming and requires a strong commitment to multi-disciplinary research in this area.

\section{Conclusion}
\label{sec-concl}
\vspace{-2ex}
After this wandering in cyberspace, have we answered the question wether cyber risk is special? As any emerging risk, the first reaction is to think it is very different from what we are used to. After reexamination, it is clear that there are commonalities with other risks that society has been dealing with rather successfully. It is why, we need to first rely on experience to manage this new risk, while understanding its specificities to adapt our responses to it. As we have seen, cyber risk is catastrophic and multi-dimensional in nature. Catastrophes, we have learned to handle them in insurance. However, the multi-dimensional aspect needs to be dealt with clear awareness, particularly for the political dimension of the psycho-cognitive layer of the cyber-space. It will require the efforts of researchers from very different fields.

Managing cyber risk involves both better understanding of that risk and better approaches to the management of that risk.  In particular, both research and practice will need to focus more on interactions: between the different layers of cyberspace; between the motivations of different actors in cyberspace; between different types of models of that risk; and between cyber security and cyber insurance. 

So, the answer to the question posed at the beginning of this survey is, as often in life, "yes and no". Yes, we need to deal with the specificities of cyber. No, we can use methods and models that have already been developed and successful for dealing with other risks, and build from them to innovate and create new efficient solutions. In any case, society as a whole should not fear cyber risk, but, fully aware of it, build up its resilience. Indeed, we will have to live with it for a long time if we want to continue benefiting from the progress IT brought to us.

We certainly have omitted some references, not being aware of them all, despite a thorough search. We apologize to the authors we did not cite.

{\bf Acknowledgments:} We would like to warmly thank Mogens Steffensen for inviting us to write this critical survey, suggesting some questions he had. 
We tried to answer them here, and used some to shape our paper.
 We are grateful to  Steven Jackson, for his careful reading of the manuscript and for pointing out to some logical conclusions. Thanks also to Simone Dalessi for having contributed to a better formulation of our text. The remaining mistakes are ours.
 
\bibliographystyle{abbrvnat}
\bibliography{Lit.bib}

\end{document}